# Designing Quality Requirements, Metrics and Indicators for Core Ontologies: Results of a Comparative Study for Process Core Ontologies

**Luis Olsina, María Fernanda Papa, Guido Tebes** and **Pablo Becker**

GIDIS_Web, Facultad de Ingeniería, UNLPam, General Pico, LP, Argentina

[olsinal, pmfer, guido_tebes, beckerp]@ing.unlpam.edu.ar

**Abstract.** This preprint specifies quality requirements for a core ontology whose ontological elements such as terms, non-taxonomic relationships, among others, are based on a foundational ontology. The quality requirements are represented in a quality model that is structured in the form of a requirements tree composed of characteristics and attributes to be measured and evaluated. An attribute represents an atomic aspect of an entity, that is, an elementary non-functional requirement that can be measured by a direct or indirect metric and evaluated by an elementary indicator. In contrast, characteristics that model less atomic aspects of an entity cannot be measured by metrics, but rather are evaluated by derived indicators generally modeled by an aggregation function. Therefore, this preprint shows the design of direct and indirect metrics in addition to the design of elementary indicators, which are used to implement measurement and evaluation activities to obtain the results of a quality requirements tree. In particular, this document shows the applicability of the designed metrics and indicators that are used by a evaluation and comparison strategy. Two process core ontologies were preselected, evaluated and compared in order to adopt strengths in the target entity named ProcessCO. The data and information resulting from this study are also recorded, as well as the outcomes of the revaluation after improvement of the target entity.

## Preprint Roadmap

This Section provides an overview of how this preprint is organized into content-specific appendices, which are supplementary material to a journal article. There are eight appendices, namely:

- Appendix I presents the quality requirements, i.e., the quality focus, characteristics and attributes to measure and evaluate the structural and reuse aspects of core ontologies. Process core ontologies are the evaluable entities in this study. Particularly, for a core ontology, we evaluate its conceptualization and associated documents that contain definitions of terms, properties, non-taxonomic relationships and axioms. Note that the concepts of Attribute, Characteristic, Non-Functional Requirement, Evaluable Entity, among others, are defined in NFRsTDO (*Non-Functional Requirements Top-Domain Ontology*) [9].
- Appendix II documents the designed direct and indirect metrics that quantify the attributes shown in Appendix I. Note that the terminology used in the template with metadata for metrics comes from MetricsLDO (*Metric-based Measurement Low-Domain Ontology*) [8]. This template is completed (instantiated) with the specific information for each metric used.
- Appendix III documents the designed elementary indicators that evaluate the attributes shown in Appendix I considering also the instantiated metrics of Appendix II accordingly. Note that the terminology used in the template with metadata for elementary indicators comes from IndicatorsLDO (*Indicator-based Evaluation Low-Domain Ontology*) [7]. This template is completed (instantiated) with the specific information for each elementary indicator used.
- Appendix IV documents the template with the metadata for data extraction, which is completed with the specific information collected from both preselected process core ontologies. Note that one of the criteria for preselection was that the core ontology should reuse or extend elements from a foundational ontology. And another criterion was that the document should contain a formal graphical representation for the conceptualization. The evaluated ontologies are: Software Process Ontology (SPO) [4] and Process Core Ontology (ProcessCO) [3]. In order to collect the data, both papers were thoroughly analyzed as well as the additional documents linked to those papers. These filled forms were the source for the measurement procedures of direct metrics that allowed us to obtain the measured values.
- Appendix V shows the measured and calculated values obtained for both ontologies using the direct and indirect metrics specified in Appendix II and the filled forms of Appendix IV.
- Appendix VI shows the values of the elementary indicators calculated from the measured values

and using the elementary function of the indicators specified in Appendix III.

- Appendix VII shows the values of the partial indicators and the global indicator calculated from the quality model specification (Appendix I) and using the LSP (*Logic Scoring of Preference*) aggregation function [5]. It is worth mentioning that the GOCAMECom (*Goal-Oriented Context-Aware Measurement, Evaluation and Comparison*) strategy [6, 12] was used for all the activities of this study. Its method and work process specifications are semantically enriched with ontologies such as those documented in [1, 2, 7, 8, 9, 10].
- Appendix VIII records the values of partial and global indicators calculated from the quality model specification (Appendix I) by using the same LSP aggregation function with the same weights and operators for both the comparative evaluation (Appendix VII) as for the revaluation of the improved entity. For the change made over ProcessCO v1.2, attribute 1.1.3 was basically considered, which implied the addition of six axioms specified in first-order logic. As a result of this change, the new version (v1.3) [2] of ProcessCO was produced.

# Appendix I: Ontological Quality Requirements Specification

*This Appendix details the designed characteristics and attributes for evaluating the structural and reuse quality of core ontologies. For a better understanding, it details first the non-functional requirements tree in which the quality evaluation focus is the Ontological Internal Quality, and then its elements (characteristics and attributes) are defined.*

1 **Ontological Internal Quality** (root characteristic of the quality evaluation focus)
- **1.1 Ontological Structural Quality**
  - *1.1.1 Defined Terms Availability*
  - *1.1.2 Defined Properties Availability*
  - *1.1.3 Formally Specified Axioms Availability*
  - **1.1.4 Balanced Relationships Availability**
    - *1.1.4.1 Defined Non-Taxonomic Relationships Availability*
    - *1.1.4.2 Balanced Non-Taxonomic / Taxonomic Relationships Availability*
- **1.2 Ontological Quality of Reuse and Compliance**
  - **1.2.1 Ontological Reuse Quality**
    - *1.2.1.1 Level of Reuse of Terms from Foundational Ontology*
    - *1.2.1.2 Level of Reuse of Non-Taxonomic Relationships from Foundational Ontology*
  - *1.2.2 Level of Use of International Standard Glossaries*

| **Characteristics /** *Attributes* | **Definition:** Degree to which… |
|---|---|
| **1. Ontological Internal Quality** (root characteristic) | … the core ontology is well structured and supports the reuse of ontological elements from a foundational ontology and adherence to standards. |
| **1.1 Ontological Structural Quality** | … the ontology is well structured as it has defined terms, defined properties, specified axioms, and is properly balanced with respect to taxonomic and non-taxonomic types of relationships. |
| *1.1.1 Defined Terms Availability* | … the ontology has its terms not only explicit but also defined in natural language. |
| *1.1.2 Defined Properties Availability* | … the ontology has its properties not only explicit but also defined in natural language. |
| *1.1.3 Formally Specified Axioms Availability* | … the ontology or associated artifact has its axioms formally specified, for example, in first-order logic. |
| **1.1.4 Balanced Relationships Availability** | … the ontology has a suitable balance between the proportion of non-taxonomic and taxonomic relationships in addition to the former are defined. |
| *1.1.4.1 Defined Non-Taxonomic Relationships Availability* | … the ontology has its non-taxonomic relationships not only explicit but also defined in natural language. |
| *1.1.4.2 Balanced Non-Taxonomic / Taxonomic Relationships Availability* | … the ontology has a balance between the size of non-taxonomic and taxonomic relationships. |
| **1.2 Ontological Quality of Reuse and Compliance** | … the core ontology supports the reuse of ontological elements from a foundational ontology and indicates adherence to international standard glossaries. |
| **1.2.1 Ontological Reuse Quality** | … the core ontology supports the reuse of ontological elements from a foundational ontology. |
| *1.2.1.1 Level of Reuse of Terms from Foundational Ontology* | … the core ontology specializes terms from a foundational ontology. |
| *1.2.1.2 Level of Reuse of Non-Taxonomic Relationships from Foundational Ontology* | … the core ontology specializes non-taxonomic relationships from a foundational ontology. |
| *1.2.2 Level of Use of International Standard Glossaries* | … the ontology uses or refers to international standard glossaries. |

# Appendix II: Specification of Indirect Metrics and their Related Direct Metrics

*In this Appendix the designed metrics are specified. For a better understanding, the indirect metric is detailed first, and then its related direct metrics. Note that only attribute 1.2.2 is quantified using a direct metric.*

## 1.1.1 Defined Terms Availability

| **Indirect Metric** |
|---|
| **Quantified Attribute name:** Defined Terms Availability<br>**Metric Name:** Percentage of Defined Terms (**%DT**) |
| **Objective:** Determine the percentage of defined terms with respect to the total of explicit terms in the ontology to be measured<br>**Author:** Luis Olsina and Pablo Becker<br>**Version:** 1.0 |
| **Calculation Procedure:**<br>Formula: **%DT = (#DT / #TT) * 100** |
| **Scale**: Numeric<br>Scale Type name: Ratio<br>Value Type: Real<br>Representation: Continuous |
| **Unit**:<br>Name: Percentage<br>Acronym:% |
| **Related Direct Metrics:**<br>#DT: Number of Defined Terms<br>#TT: Total Number of Terms |

| **Direct Metric** |
|---|
| **Quantified Attribute name:** Size of defined Terms<br>**Metric Name:** Number of Defined Terms (**#DT**) |
| **Objective:** Determine the number of defined terms in the ontology to be measured<br>**Author:** Luis Olsina and Pablo Becker<br>**Version:** 1.0 |
| **Measurement Procedure:**<br>Type: Objective<br>Specification: |
| #DT= 0<br>For each defined term of the ontology in a given natural language, do: #DT++ |
| **Scale**: Numeric<br>Scale Type name: Absolute<br>Value Type: Integer<br>Representation: Discrete |
| **Unit**:<br>Name: Term<br>Description: a term of an ontology represents an Entity (with the semantics of Thing according to ThingFO [10, 11]), an Entity Category or an Assertion<br>Acronym: T |

| **Direct Metric** |
|---|
| **Quantified Attribute name:** Total size of Terms<br>**Metric Name:** Total Number of Terms (**#TT**) |
| **Objective:** Determine the total number of terms in the ontology to be measured<br>**Author:** Luis Olsina and Pablo Becker<br>**Version:** 1.0 |
| **Measurement Procedure:**<br>Type: Objective<br>Specification:<br>#TT= 0<br>For each explicit ontology term in a given natural language and regardless of whether it is defined or not, do: #TT++ |
| **Scale**: Numeric<br>Scale Type name: Absolute<br>Value Type: Integer<br>Representation: Discrete |
| **Unit**:<br>Name: Term<br>Description: a term of an ontology represents an Entity (with the semantics of Thing according to ThingFO [10, 11]), an Entity Category or an Assertion<br>Acronym: T |

### 1.1.2 Defined Properties Availability

| **Indirect Metric** |
|---|
| **Quantified Attribute name:** Defined Properties Availability<br>**Metric Name:** Percentage of Defined Properties (**%DP**) |
| **Objective:** Determine the percentage of defined properties with respect to the total of explicit properties in the ontology to be measured<br>**Author:** Luis Olsina and Pablo Becker<br>**Version:** 1.0 |
| **Calculation Procedure:**<br>Formula: **%DP = (#DP / #TP) * 100**; If **#TP** = 0, then **%DP** = 0 |
| **Scale**: Numeric<br>Scale Type name: Ratio<br>Value Type: Real<br>Representation: Continuous |
| **Unit**:<br>Name: Percentage<br>Acronym: % |
| **Related Direct Metrics:**<br>#DP: Number of Defined Properties<br>#TP: Total Number of Properties |

| **Direct Metric** |
|---|
| **Quantified Attribute name:** Size of defined Properties<br>**Metric Name:** Number of Defined Properties (**#DP**) |
| **Objective:** Determine the number of defined properties in the ontology to be measured<br>**Author:** Luis Olsina and Pablo Becker<br>**Version:** 1.0 |
| **Measurement Procedure:**<br>Type: Objective<br>Specification:<br>#DP= 0<br>For each defined property of the ontology in a given natural language, do: #DP++ |
| **Scale**: Numeric<br>Scale Type name: Absolute<br>Value Type: Integer<br>Representation: Discrete |
| **Unit**:<br>Name: Property<br>Description: A property (with the semantics of Property according to ThingFO [10, 11]) represents the structure or parts of an entity or term of an ontology<br>Acronym: P |

| **Direct Metric** |
|---|
| **Quantified Attribute name:** Total size of Properties<br>**Metric Name:** Total Number of Properties (**#TP**) |
| **Objective:** Determine the total number of properties in the ontology to be measured<br>**Author:** Luis Olsina and Pablo Becker<br>**Version:** 1.0 |
| **Measurement Procedure:**<br>Type: Objective<br>Specification:<br>#TP= 0<br>For each explicit ontology property in a given natural language and regardless of whether it is defined or not, do: #TP++ |
| **Scale**: Numeric<br>Scale Type name: Absolute<br>Value Type: Integer<br>Representation: Discrete |
| **Unit**:<br>Name: Property<br>Description: A property (with the semantics of Property according to ThingFO [10, 11]) represents the structure or parts of an entity or term of an ontology<br>Acronym: P |

### 1.1.3 Formally Specified Axioms Availability

| **Indirect Metric** |
|---|
| **Quantified Attribute name:** Formally Specified Axioms Availability<br>**Metric Name:** Percentage of Specified Axioms **(%SA)** |
| **Objective:** Determine the percentage of formally specified axioms with respect to the total of axioms described in the ontology or related artifact to be measured<br>**Author:** Luis Olsina and Pablo Becker<br>**Version:** 1.0 |
| **Calculation Procedure:**<br>Formula: **%SA = (#SA / #TA) * 100**; If **#TA** = 0, then **%SA** = 0 |
| **Scale**: Numeric<br>Scale Type name: Ratio<br>Value Type: Real<br>Representation: Continuous |
| **Unit**: |
| Name: Percentage<br>Acronym: % |
| **Related Direct Metrics:**<br>#SA: Number of Specified Axioms<br>#TA: Total Number of Axioms |

| **Direct Metric** |
|---|
| **Quantified Attribute name:** Size of specified Axioms<br>**Metric Name:** Number of Specified Axioms (**#SA**) |
| **Objective:** Determine the number of formally specified axioms (e.g., in first-order logic) in the ontology or related artifact to be measured<br>**Author:** Luis Olsina and Pablo Becker<br>**Version:** 1.0 |
| **Measurement Procedure:**<br>Type: Objective<br>Specification:<br>#SA = 0<br>For each formally specified axioms, do: #SA++ |
| **Scale**: Numeric<br>Scale Type name: Absolute<br>Value Type: Integer<br>Representation: Discrete |
| **Unit**:<br>Name: Axiom<br>Description: An axiom (with the semantics of Constraint-related Assertion according to ThingFO [10, 11]) represents a constraint of some ontology elements<br>Acronym: A |

| **Direct Metric** |
|---|
| **Quantified Attribute name:** Total size of Axioms<br>**Metric Name:** Total Number of Axioms (**#TA**) |
| **Objective:** Determine the total number of axioms described in the ontology or related artifact to be measured<br>**Author:** Luis Olsina and Pablo Becker<br>**Version:** 1.0 |
| **Measurement Procedure:**<br>Type: Objective<br>Specification: |
| #TA= 0<br>For each axiom described in the ontology in a given natural language, and regardless of whether it is formally specified or not, do: #TA++ |
| **Scale**: Numeric<br>Scale Type name: Absolute<br>Value Type: Integer<br>Representation: Discrete |
| **Unit**:<br>Name: Axiom<br>Description: An axiom (with the semantics of Constraint-related Assertion according to ThingFO [10, 11]) represents a constraint of some ontology elements<br>Acronym: A |

## 1.1.4.1 Defined Non-Taxonomic Relationships Availability

| **Indirect Metric** |
|---|
| **Quantified Attribute name:** Defined Non-Taxonomic Relationships Availability<br>**Metric Name:** Percentage of Defined Non-Taxonomic Relationships (**%DNTR**) |
| **Objective:** Determine the percentage of defined non-taxonomic relationships with respect to the total of explicit non-taxonomic relationships in the ontology to be measured<br>**Author:** Luis Olsina and Pablo Becker<br>**Version:** 1.0 |
| **Calculation Procedure:**<br>Formula: **%DNTR = (#DNTR / #TNTR) * 100**; If **#TNTR** = 0, then **%DNTR** = 0 |
| **Scale**: Numeric<br>Scale Type name: Ratio<br>Value Type: Real<br>Representation: Continuous |
| **Unit**:<br>Name: Percentage<br>Acronym: % |
| **Related Direct Metrics:**<br>#DNTR: Number of Defined Non-Taxonomic Relationships<br>#TNTR: Total Number of Non-Taxonomic Relationships |

| **Direct Metric** |
|---|
| **Quantified Attribute name:** Size of defined Non-Taxonomic Relationships<br>**Metric Name:** Number of Defined Non-Taxonomic Relationships (**#DNTR**) |
| **Objective:** Determine the number of defined non-taxonomic relationships in the ontology to be measured<br>**Author:** Luis Olsina and Pablo Becker<br>**Version:** 1.0 |
| **Measurement Procedure:**<br>Type: Objective<br>Specification:<br>#DNTR = 0<br>For each defined non-taxonomic relationship in the ontology in a given natural language, do: #DNTR++. Note: a non-taxonomic relationship is one that links two terms of the ontology but does not belong to the group of *kind_of* (or *is_a*) or *whole-part* (or *part_of*) relationships |
| **Scale**: Numeric<br>Scale Type name: Absolute<br>Value Type: Integer<br>Representation: Discrete |
| **Unit**:<br>Name: Relationship<br>Description: It represents the way in which the terms of an ontology are linked or related<br>Acronym: R |

| **Direct Metric** |
|---|
| **Quantified Attribute name:** Size of Non-Taxonomic Relationships<br>**Metric Name:** Total Number of Non-Taxonomic Relationships (**#TNTR**) |
| **Objective:** Determine the total number of non-taxonomic relationships which are represented in the ontology to be measured<br>**Author:** Luis Olsina and Pablo Becker<br>**Version:** 1.0 |
| **Measurement Procedure:**<br>Type: Objective<br>Specification:<br>#TNTR= 0<br>For each explicit non-taxonomic relationship of the ontology in a given natural language, and regardless of whether it is defined or not, do: #TNTR ++. Note: a non-taxonomic relationship is one that links two terms of the ontology but does not belong to the group of *kind_of* (or *is_a*) or *whole-part* (or *part_of*) relationships |
| **Scale**: Numeric<br>Scale Type name: Absolute<br>Value Type: Integer<br>Representation: Discrete |
| **Unit**: |
| Name: Relationship<br>Description: It represents the way in which the terms of an ontology are linked or related<br>Acronym: R |

#### 1.1.4.2 Balanced Non-Taxonomic / Taxonomic Relationships Availability

| **Indirect Metric** |
|---|
| **Quantified Attribute name:** Balanced Non-Taxonomic / Taxonomic Relationships Availability<br>**Metric Name:** Percentage of Balanced Non-Taxonomic Relationships (**%BNTR**) |
| **Objective:** Determine the percentage of non-taxonomic relationships with respect to the total of relationships represented in the ontology to be measured<br>**Author:** Luis Olsina and Pablo Becker<br>**Version:** 1.0 |
| **Calculation Procedure:**<br>Formula: **%BNTR = (#TNTR / #TR) * 100** |
| **Scale**: Numeric<br>Scale Type name: Ratio<br>Value Type: Real<br>Representation: Continuous |
| **Unit**:<br>Name: Percentage<br>Acronym: % |
| **Related Direct Metrics:**<br>#TNTR: Total Number of Non-Taxonomic Relationships<br>#TR: Total Number of Relationships |

| **Direct Metric** |
|---|
| **Quantified Attribute name:** Size of Relationships<br>**Metric Name:** Total Number of Relationships (#TR) |
| **Objective:** Determine the total number of relationships represented in the ontology to be measured, regardless of whether they are taxonomic or not<br>**Author:** Luis Olsina and Pablo Becker<br>**Version:** 1.0 |
| **Measurement Procedure:**<br>Type: Objective<br>Specification:<br>#TR= 0 |
| For each relationship represented in the ontology, regardless of whether it is taxonomic or not, do: #TR++ |
| **Scale**: Numeric<br>Scale Type name: Absolute<br>Value Type: Integer<br>Representation: Discrete |
| **Unit**:<br>Name: Relationship<br>Description: It represents the way in which the terms of an ontology are linked or related<br>Acronym: R |

#### 1.2.1.1 Level of Reuse of Terms from Foundational Ontology

| |
|---|
| **Indirect Metric** |
| **Quantified Attribute name:** Level of Reuse of Terms from Foundational Ontology<br>**Metric Name:** Percentage of Specialized Terms from Foundational Ontology (**%STFO**) |
| **Objective:** Determine in the core ontology to be measured the percentage of specialized terms from a foundational ontology either directly (e.g., using stereotypes, among others reuse mechanism as inheritance), or indirectly through a core ontology which extends from the foundational.<br>**Author:** Luis Olsina and Pablo Becker<br>**Version:** 1.0 |
| **Calculation Procedure:**<br>Formula: **%STFO = ((#STDFO + #STIFO) / #TT) * 100** |
| **Scale**: Numeric<br>Scale Type name: Ratio<br>Value Type: Real<br>Representation: Continuous |
| **Unit**:<br>Name: Percentage<br>Acronym: % |
| **Related Direct Metrics:**<br>#STDFO: Number of Specialized Terms Directly from Foundational Ontology<br>#STIFO: Number of Specialized Terms Indirectly from Foundational Ontology<br>#TT: Total Number of Terms |
| **Direct Metric** |
| **Quantified Attribute name:** Size of Specialized Terms<br>**Metric Name:** Number of Specialized Terms Directly from Foundational Ontology (**#STDFO**) |
| **Objective:** Determine in the core ontology to be measured the number of specialized terms directly from a foundational ontology (e.g., using stereotypes, among others reuse mechanism as inheritance)<br>**Author:** Luis Olsina and Pablo Becker<br>**Version:** 1.0 |
| **Measurement Procedure:**<br>Type: Objective<br>Specification:<br>#STDFO = 0<br>For each term in the core ontology to be measured which is specialized directly from a term belonging to a foundational ontology, do: #STDFO++ |
| **Scale**: Numeric<br>Scale Type name: Absolute<br>Value Type: Integer<br>Representation: Discrete |
| **Unit**:<br>Name: Term<br>Description: a term of an ontology represents an Entity (with the semantics of Thing according to ThingFO [10, 11]), an Entity Category or an Assertion<br>Acronym: T |

| **Direct Metric** |
|---|
| **Quantified Attribute name:** Size of Specialized Terms<br>**Metric Name:** Number of Specialized Terms Indirectly from Foundational Ontology (**#STIFO**) |
| **Objective:** Determine in the core ontology to be measured the number of specialized terms indirectly (e.g., using stereotypes, among others reuse mechanism as inheritance), that is, terms from another core ontology that in turn reuses the term from the foundational ontology<br>**Author:** Luis Olsina and Pablo Becker<br>**Version:** 1.0 |
| **Measurement Procedure:**<br>Type: Objective<br>Specification:<br>#STIFO = 0<br>For each term in the core ontology to be measured which is specialized indirectly, that is, from a term belonging to another core ontology that in turn reuses the term from the foundational ontology, do: #STIFO++ |
| **Scale**: Numeric<br>Scale Type name: Absolute<br>Value Type: Integer<br>Representation: Discrete |
| **Unit**: |
| Name: Term<br>Description: a term of an ontology, represents an Entity (with the semantics of Thing according to ThingFO [10, 11]), an Entity Category or an Assertion<br>Acronym: T |

#### 1.2.1.2 Level of Reuse of Non-Taxonomic Relationships from Foundational Ontology

| **Indirect Metric** |
|---|
| **Quantified Attribute name:** Level of Reuse of Non-Taxonomic Relationships from Foundational Ontology<br>**Metric Name:** Percentage of Specialized Non-Taxonomic Relationships from Foundational Ontology (**%SNTRFO**) |
| **Objective:** Determine the percentage of specialized non-taxonomic relationships from a foundational ontology with respect to the total of non-taxonomic relationships explicit in the ontology to be measured<br>**Author:** Luis Olsina and Pablo Becker<br>**Version:** 1.0 |
| **Calculation Procedure:**<br>Formula: **%SNTRFO = (#SNTRFO / #TNTR) * 100** ; If **#TNTR** = 0, then **%SNTRFO** = 0 |
| **Scale**: Numeric<br>Scale Type name: Ratio<br>Value Type: Real<br>Representation: Continuous |
| **Unit**:<br>Name: Percentage<br>Acronym: % |
| **Related Direct Metrics:**<br>#SNTRFO: Number of Specialized Non-Taxonomic Relationships from Foundational Ontology<br>#TNTR: Total Number of Non-Taxonomic Relationships |

| **Direct Metric** |
|---|
| **Quantified Attribute name:** Size of Specialized Non-Taxonomic Relationships<br>**Metric Name:** Number of Specialized Non-Taxonomic Relationships from Foundational Ontology (**#SNTRFO**) |
| **Objective:** Determine in the ontology to be measured the number of specialized non-taxonomic relationships from a foundational ontology<br>**Author:** Luis Olsina and Pablo Becker<br>**Version:** 1.0 |
| **Measurement Procedure:**<br>Type: Objective<br>Specification:<br>#SNTRFO = 0<br>For each non-taxonomic relationship in the ontology to be measured which is specialized from a non-taxonomic relationship belonging to a foundational ontology, do: #SNTRFO++. Note 1: a non-taxonomic relationship is one that links two terms of the ontology but does not belong to the group of *kind_of* (or *is_a*) or *whole-part* (or *part_of*) relationships. Note 2: a non-taxonomic relationship in a given ontology could be specialized from a higher level ontology. It could be specialized as a subset or as a refinement by redefinition. |
| **Scale**: Numeric<br>Scale Type name: Absolute<br>Value Type: Integer<br>Representation: Discrete |
| **Unit**:<br>Name: Relationship<br>Description: It represents the way in which the terms of an ontology are linked or related<br>Acronym: R |

### 1.2.2 Level of Use of International Standard Glossaries

| **Direct Metric** |
|---|
| **Quantified Attribute name:** Level of Use of International Standard Glossaries<br>**Metric Name:** Number of Used International Standard Glossaries (#UISG) |
| **Objective:** Determine the number of international standard glossaries used or referred by the ontology to be measured.<br>**Author:** Guido Tebes and Luis Olsina<br>**Version:** 1.0 |
| **Measurement Procedure:**<br>Type: Objective<br>Specification:<br>#UISG = 0<br>For each international standard glossary used or referred to by the ontology, do: #UISG++ |
| **Scale**: Numerical<br>Scale Type name: Absolute<br>Value Type: Integer<br>Representation: Discrete |
| **Unit**:<br>Name: International Standard Glossary<br>Description: it is an international standard glossary that can be an official or de-facto standard.<br>Acronym: ISG |

# Appendix III: Elementary Indicators Specification

*In this Appendix the designed elementary indicators that evaluate the attributes are specified.*

## 1.1.1 Defined Terms Availability

| **Elementary Indicator** |
|---|
| **Interpreted Attribute name:** Defined Terms Availability<br>**Instantiated Metric name:** Porcentage of Defined Terms (**%DT**)<br>**Indicator name:** Performance Level in Defined Terms Availability (**PL_DTA**) |
| **Author:** Luis Olsina and Pablo Becker<br>**Version:** 1.0 |
| **Elementary Function:**<br>Specification:<br>**PL_DTA** = **%DT**<br><br>**Decision Criteria** [Acceptability Levels]<br>Name: **Unsatisfactory**<br>Range: [0; 60]<br>Description: Indicates that corrective actions must be performed with high priority<br>Name: **Marginal**<br>Range: (60; 85]<br>Description: Indicates that corrective actions should be performed<br>Name: **Satisfactory**<br>Range: (85; 100]<br>Description: Indicates that corrective actions are not necessary since the attribute meets the required quality satisfaction level |
| **Scale:** Numeric<br>Scale Type name: Ratio<br>Value Type: Real<br>Representation: Continuous |
| **Unit:**<br>Name: Percentage<br>Acronym: % |

## 1.1.2 Defined Properties Availability

| **Elementary Indicator** |
|---|
| **Interpreted Attribute name:** Defined Properties Availability<br>**Instantiated Metric name:** Percentage of Defined Properties (**%DP**)<br>**Indicator name:** Performance Level in Defined Properties Availability (**PL_DPA**) |

| | |
|---|---|
| **Author:** Luis Olsina and Pablo Becker | **Version:** 1.0 |
| **Elementary Function:**<br>Specification:<br>**PL_DPA = %DP**<br><br>**Decision Criteria** [Acceptability Levels]<br>Name: **Unsatisfactory**<br>Range: [0; 60]<br>Description: Indicates that corrective actions must be performed with high priority<br>Name: **Marginal**<br>Range: (60; 85]<br>Description: Indicates that corrective actions should be performed<br>Name: **Satisfactory**<br>Range: (85; 100]<br>Description: Indicates that corrective actions are not necessary since the attribute meets the required quality satisfaction level | |
| **Scale:** Numeric<br>Scale Type name: Ratio<br>Value Type: Real<br>Representation: Continuous | |
| **Unit:**<br>Name: Percentage<br>Acronym: % | |

## 1.1.3 Formally Specified Axioms Availability

| **Elementary Indicator** | |
|---|---|
| **Interpreted Attribute name:** Formally Specified Axioms Availability<br>**Instantiated Metric name:** Percentage of Specified Axioms **(%SA)**<br>**Indicator name:** Performance Level in Formally Specified Axioms Availability (**PL_FSAA**) | |
| **Author:** Luis Olsina and Pablo Becker | **Version:** 1.0 |
| **Elementary Function:**<br>Specification:<br>**PL_FSAA = %SA**<br><br>**Decision Criteria** [Acceptability Levels]<br>Name: **Unsatisfactory**<br>Range: [0; 60]<br>Description: Indicates that corrective actions must be performed with high priority<br>Name: **Marginal**<br>Range: (60; 85]<br>Description: Indicates that corrective actions should be performed | |

| |
|---|
| Name: **Satisfactory**<br>Range: (85; 100]<br>Description: Indicates that corrective actions are not necessary since the attribute meets the required quality satisfaction level |
| **Scale:** Numeric<br>Scale Type name: Ratio<br>Value Type: Real<br>Representation: Continuous |
| **Unit:**<br>Name: Percentage<br>Acronym: % |

### 1.1.4.1 Defined Non-Taxonomic Relationships Availability

| **Elementary Indicator** |
|---|
| **Interpreted Attribute name:** Defined Non-Taxonomic Relationships Availability<br>**Instantiated Metric name:** Percentage of Defined Non-Taxonomic Relationships (**%DNTR**)<br>**Indicator name:** Performance Level in Defined Non-Taxonomic Relationships Availability (**PL_DNTRA**) |
| **Author:** Luis Olsina and Pablo Becker **Version:** 1.0 |
| **Elementary Function:**<br>Specification:<br>**PL_DNTRA = %DNTR**<br><br>**Decision Criteria** [Acceptability Levels]<br>Name: **Unsatisfactory**<br>Range: [0; 60]<br>Description: Indicates that corrective actions must be performed with high priority<br>Name: **Marginal**<br>Range: (60; 85]<br>Description: Indicates that corrective actions should be performed<br>Name: **Satisfactory**<br>Range: (85; 100]<br>Description: Indicates that corrective actions are not necessary since the attribute meets the required quality satisfaction level |
| **Scale:** Numeric<br>Scale Type name: Ratio<br>Value Type: Real<br>Representation: Continuous |
| **Unit:**<br>Name: Percentage |
| Acronym: % |

### 1.1.4.2 Balanced Non-Taxonomic / Taxonomic Relationships Availability

| **Elementary Indicator** | |
|---|---|
| **Interpreted Attribute name:** Balanced Non-Taxonomic / Taxonomic Relationships Availability<br>**Instantiated Metric name:** Percentage of Balanced Non-Taxonomic Relationships (**%BNTR**)<br>**Indicator name:** Performance Level in Balanced Non-Taxonomic / Taxonomic Relationships Availability (**PL_BNTTRA**) | |
| **Author:** Pablo Becker and Luis Olsina | **Version:** 1.0 |

**Elementary Function:**
Specification:

PL_BNTTRA (*x*)

$$\text{PL_BNTTRA}(x)= \begin{cases} 6x; & 0 \le x < 10 \quad (1) \\ \frac{5}{6}x + \frac{155}{3}; & 10 \le x < 40 \quad (2) \\ \frac{3}{2}x + 25; & 40 \le x \le 50 \quad (3) \\ -\frac{3}{2}x + 175; & 50 < x \le 60 \quad (4) \\ -\frac{5}{6}x + 135; & 60 < x \le 90 \quad (5) \\ -6x + 600; & 90 < x \le 100 \quad (6) \end{cases}$$

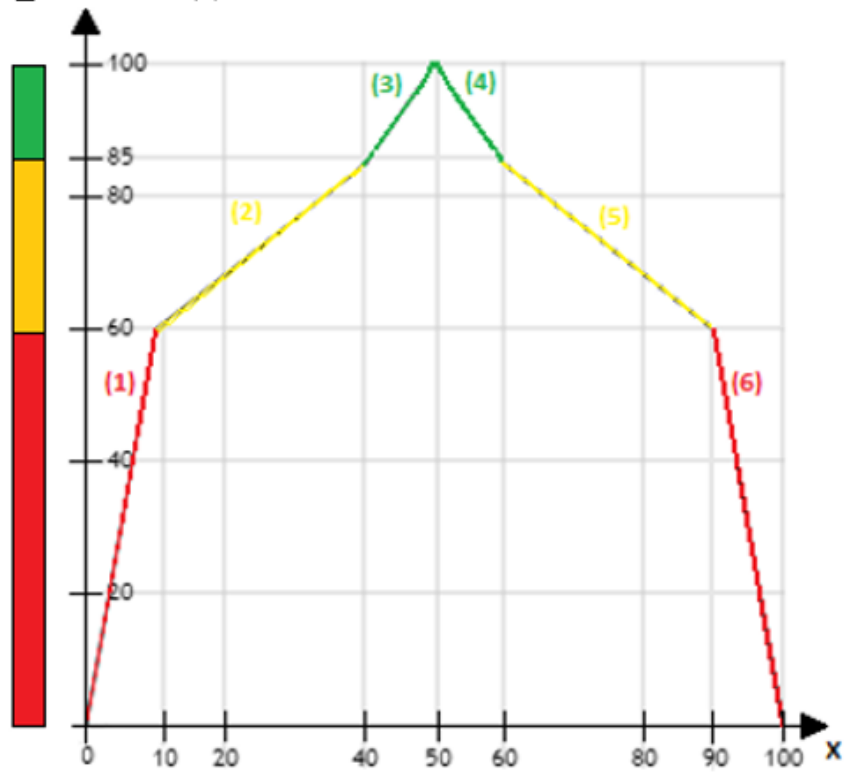

where *x* is the *%BNTR* metric

**Decision Criteria** [Acceptability Levels]
Name: **Unsatisfactory**
Range: [0; 60]
Description: Indicates that corrective actions must be performed with high priority
Name: **Marginal**
Range: (60; 85]
Description: Indicates that corrective actions should be performed
Name: **Satisfactory**
Range: (85; 100]
Description: Indicates that corrective actions are not necessary since the attribute meets the required quality satisfaction level

**Scale:** Numeric
Scale Type name: Ratio
Value Type: Real
Representation: Continuous

**Unit:**

Name: Percentage
Acronym: %

#### 1.2.1.1 Level of Reuse of Terms from Foundational Ontology

**Elementary Indicator**

**Interpreted Attribute name:** Level of Reuse of Terms from Foundational Ontology
**Instantiated Metric name:** Percentage of Specialized Terms from Foundational Ontology (**%STFO**)
**Indicator name:** Performance in the Level of Reuse of Terms from Foundational Ontology (**P_LRTFO**)

**Author:** Pablo Becker and Luis Olsina **Version:** 1.0

**Elementary Function:**
Specification:

$$\mathbf{P_LRTFO} = \begin{cases} 100; & \%\mathbf{STFO} = 100 \\ 85; & 95 \leq \%\mathbf{STFO} < 100 \\ 60; & 70 \leq \%\mathbf{STFO} < 95 \\ 0; & 0 \leq \%\mathbf{STFO} < 70 \end{cases}$$

**Decision Criteria** [Acceptability Levels]
Name: **Unsatisfactory**
Range: [0; 60]
Description: Indicates that corrective actions must be performed with high priority
Name: **Marginal**
Range: (60; 85]
Description: Indicates that corrective actions should be performed
Name: **Satisfactory**
Range: (85; 100]
Description: Indicates that corrective actions are not necessary since the attribute meets the required quality satisfaction level

**Scale:** Numeric
Scale Type name: Ratio
Value Type: Real
Representation: Continuous

**Unit:**
Name: Percentage
Acronym: %

#### 1.2.1.2 Level of Reuse of Non-Taxonomic Relationships from Foundational Ontology

**Elementary Indicator**

| |
|---|
| **Interpreted Attribute name:** Level of Reuse of Non-Taxonomic Relationships from Foundational Ontology<br>**Instantiated Metric name:** Percentage of Specialized Non-Taxonomic Relationships from Foundational Ontology (**%SNTRFO**)<br>**Indicator name:** Performance in the Level of Reuse of Non-Taxonomic Relationships from Foundational Ontology (**P_LRNTRFO**) |
| **Author:** Pablo Becker and Luis Olsina **Version:** 1.0 |
| **Elementary Function:**<br>Specification:<br>$$\mathbf{P_LRNTRFO} = \begin{cases} 100; & \mathbf{\%SNTRFO} = 100 \\ 85; & 95 \leq \mathbf{\%SNTRFO} < 100 \\ 60; & 70 \leq \mathbf{\%SNTRFO} < 95 \\ 20; & 20 \leq \mathbf{\%SNTRFO} < 70 \\ 0; & 0 \leq \mathbf{\%SNTRFO} < 20 \end{cases}$$<br>**Decision Criteria** [Acceptability Levels]<br>Name: **Unsatisfactory**<br>Range: [0; 60]<br>Description: Indicates that corrective actions must be performed with high priority<br>Name: **Marginal**<br>Range: (60; 85]<br>Description: Indicates that corrective actions should be performed<br>Name: **Satisfactory**<br>Range: (85; 100]<br>Description: Indicates that corrective actions are not necessary since the attribute meets the required quality satisfaction level |
| **Scale:** Numeric<br>Scale Type name: Ratio<br>Value Type: Real<br>Representation: Continuous |
| **Unit:**<br>Name: Percentage<br>Acronym: % |

## 1.2.2 Level of Use of International Standard Glossaries

| **Elementary Indicator** |
|---|
| **Interpreted Attribute name:** Level of Use of International Standard Glossaries<br>**Instantiated Metric name:** Number of Used International Standard Glossaries (**#UISG**)<br>**Indicator name:** Performance in the Level of Use of International Standard Glossaries (**P_LUISG)** |
| **Author:** Guido Tebes and Luis Olsina **Version:** 1.0 |
| **Elementary Function:** |
| Specification:<br>$$\mathbf{P_LUISG} = \begin{cases} 0\ if\ \ \mathbf{UISG} = 0 \\ 75\ if\ \ \mathbf{UISG} = 1 \\ 100\ if\ \mathbf{UISG} \geq 2 \end{cases}$$<br><br>**Decision Criteria** [Acceptability Levels]<br>Name: **Unsatisfactory**<br>Range: [0; 60]<br>Description: Indicates that corrective actions must be performed with high priority<br>Name: **Marginal**<br>Range: (60; 85]<br>Description: Indicates that corrective actions should be performed<br>Name: **Satisfactory**<br>Range: (85; 100]<br>Description: Indicates that corrective actions are not necessary since the attribute meets the required quality satisfaction level |
| **Scale:** Numeric<br>Scale Type name: Ratio<br>Value Type: Real<br>Representation: Continuous |
| **Unit:**<br>Name: Percentage<br>Acronym: % |

# Appendix IV: Data collection

*This Appendix instantiates the data collection template for the two Process Core Ontologies to be evaluated and compared, namely: SPO [4] and ProcessCO [3].*

| | |
|---|---|
| **Collector name** | Maria Fernanda Papa and Luis Olsina |
| **Article title** | Using a foundational ontology for reengineering a software process ontology. |
| **Author/s of the article** | Bringuente, Ana C. O.<br>Falbo, Ricardo A.<br>Guizzardi, Giancarlo |
| **References** | de Oliveira Bringuente, A. C., de Almeida Falbo, R., & Guizzardi, G. (2011). Using a foundational ontology for reengineering a software process ontology. Journal of Information and Data Management, 2(3), 511-511 (see reference [4])<br><br>Note that the Terms explicitly mentioned in [4], their definitions were collected from the following URL (last updated in 2017, http://dev.nemo.inf.ufes.br/seon/SPO.html), when available. |
| **Name of the preselected ontology** | Software Process Ontology (SPO) |
| **Specified concepts used to describe the process core ontology** | **Terms**:<br>1. **Activity Occurrence** denotes particular actions that take place in specific time intervals caused by scheduled activities.<br>2. **Atomic Activity Occurrence**<br>3. **Atomic Project Activity**<br>4. **Atomic Standard Activity** (synonym in 2017: **Simple Standard Activity** is a standard activity that is not further decomposed into other standard activities.)<br>5. **Complex Activity Occurrence**<br>6. **Complex Project Activity**<br>7. **Complex Standard Activity** (synonym in 2017: **Composite Standard Activity** is a Standard Activity composed of other Standard Activities.)<br>8. **General Project Process** is a project process that is composed of specific project processes, allowing an organization to define sub-processes that are part of a general project process. |
| | 9. **General Standard Process** (in 2017) is a standard process that is composed of specific standard processes, allowing an organization to establish sub-processes that are part of a complete standard process.<br>10. **Hardware Equipment** (in 2017) is physical object used for running software programs or to support some related action.<br>11. **Hardware Resource** (in 2017) is a hardware equipment when used as resource of some process activity.<br>12. **Hardware Resource Participation** (in 2017) is the participation of a hardware equipment as a resource in a performed activity.<br>13. **Hardware Type** (in 2017) is an object kind which is the powertype of hardware equipment, classifying its specializations.<br>14. **Human Resource** is a type of Agent that can be allocated to the project activities.<br>15. **Human Resource Allocation**<br>16. **Human Resource Participation** is the participation of a specific human resource in a complex activity occurrence.<br>17. **Human Role** is a social role in UFO, defined by a normative description (in the case, a job plan) recognized by the organization.<br>18. **Job Plan**<br>19. **Person** is a type of Human Agent, which is assigned to a project team for performing certain human roles.<br>20. **Process Occurrence(s)** are complex actions caused by scheduled Process that take place in specific time intervals.<br>21. **Project Activity** (in 2017) is an intended activity defined to be performed within a project, and thus that is part of a project process.<br>22. **Project Process** (in 2017) is an intended process defined to be performed within a project.<br>23. **Project Team** is a type of collective social agent that can allocate human resources to perform scheduled activities.<br>24. **Project Team Allocation** is the allocation of a person to a project team, which is composed of a commitment of the person |
| | to the team in performing according to the human roles assigned to her/him, and a claim of the team towards the person.<br>25. **Resource** (in 2017) is a software product or hardware equipment when used by a process activity<br>26. **Scheduled Activity** (in 2017) is an intended activity defining the time interval it is planned to be performed.<br>27. **Scheduled Process** (in 2017) is an intended process defining the time interval it is planned to be performed.<br>28. **Software Product** (in 2017) is one or more computer programs together with any accompanying auxiliary items, such as documentation, delivered under a single name, ready for use.<br>29. **Software Product Type** (in 2017) is an artifact type which is the powertype of software product, classifying its specializations.<br>30. **Software Resource** (in 2017) is a software product when used as resource of some process activity.<br>31. **Software Resource Participation** (in 2017) is the participation of a software product as resource in a performed activity.<br>32. **Specific Project Process**<br>33. **Specific Standard Process is** (in 2017) a standard process that can be part of a general standard process of a project or organization, necessarily composed of at least two standard activities.<br>34. **Standard Activity** (in 2017) is an action universal representing a generic activity institutionalized as part of a standard process in an organization, establishing general information such as dependence on other activities, types of artifacts created, changed and used, required software and hardware resources, adopted procedures, and roles to perform it.<br>35. **Standard Process** (in 2017) is a complex action universal representing a generic process institutionalized in an organization, establishing basic requirements for Intended processes to be performed in that organization or in its projects.<br>36. **Standard Process Definition Document** (in 2017) is a plan description, recognized by an organization, which describes a set of standard processes.<br><br>**<u>Attributes (Properties)</u>** are not available in [4]. |
| | **<u>Relationships</u>**:<br>**Taxonomic:**<br>**part_of (TermX_whole, TermY_part)**<br>1. part_of (General Standard Process, Specific Standard Process)<br>2. part_of (Specific Standard Process, Standard Activity)<br>3. part_of (Complex Standard Activity, Standard Activity)<br>4. part_of (General Project Process, Specific Project Process)<br>5. part_of (Specific Project Process, Project Activity)<br>6. part_of (Process Occurrence, Activity Occurrence)<br>7. part_of (Complex Project Activity, Project Activity)<br>8. part_of (Complex Activity Occurrence, Activity Occurrence)<br>**is_a (TermX_type, TermY_subtype)**<br>9. is_a (Standard Process, General Standard Process)<br>10. is_a (Standard Process, Specific Standard Process)<br>11. is_a (Project Process, General Project Process)<br>12. is_a (Project Process, Specific Project Process)<br>13. is_a (Project Process, Scheduled Process)<br>14. is_a (Standard Activity, Complex Standard Activity)<br>15. is_a (Standard Activity, Atomic Standard Activity)<br>16. is_a (Project Activity, Scheduled Activity)<br>17. is_a (Project Activity, Atomic Project Activity)<br>18. is_a (Project Activity, Complex Project Activity)<br>19. is_a (Activity Occurrence, Complex Activity Occurrence)<br>20. is_a (Activity Occurrence, Atomic Activity Occurrence)<br>21. is_a (Atomic Activity Occurrence, Human Resource Participation)<br>22. is_a (Person, Human Resource)<br>23. is_a (Resource, Hardware Resource) |
| | 24. is_a (Resource, Software Resource)<br>25. is_a (Hardware Equipment, Hardware Resource)<br>26. is_a (Software Product, Software Resource)<br><br>**Non-Taxonomic:**<br><br>**Relation_name (TermX, TermY)**<br>1. **allocates** (Project Team Allocation, Human Resource)<br>2. **caused by** (Process Occurrence, Scheduled Process): Process occurrences are caused by scheduled processes.<br>3. **caused by** (Activity Occurrence, Scheduled Activity): Activity occurrences are caused by scheduled activities.<br>4. **defines** (Job Plan, Human Role)<br>5. **depends on** (Standard Activity, Standard Activity)<br>6. **depends on** (Activity Occurrence, Activity Occurrence):<br>7. **depends on** (Project Activity, Project Activity)<br>8. **describes** (Standard Process Definition Document, Standard Process): standard process definition document as the plan description that describes the standard process.<br>9. **existential dependence of** (Human Resource Allocation, Project Team Allocation): for a human resource allocation to exist (for a human resource be allocated to a scheduled activity), she/he has to be first allocated to the project team.<br>10. **has** (Project Team, Project Team Allocation)<br>11. **is to be performed by** (Standard Activity, Human Role)<br>12. **is to be performed by** (Project Activity, Human Role)<br>13. **is to be performed using** (Standard Activity, Hardware Type)<br>14. **is to be performed using** (Standard Activity, Software Product Type)<br>15. **is to be performed using** (Project Activity, Hardware Type)<br>16. **is to be performed using** (Project Activity, Software Product Type) |
| | 17. **is to perform** (Project Team Allocation, Human Role): a project team allocation is composed of a commitment of the person to the team in performing according to the human roles assigned to her/him, and a claim of the team towards the person.<br>18. **is to perform** (Human Resource Allocation, Scheduled Activity): we can allocate human resources to perform scheduled activities<br>19. **participation of** (Human Resource Participation, Human Resource): It is the participation of a specific human resource in a complex activity occurrence.<br>20. **participation of** (Software Resource Participation, Software Resource)<br>21. **participation of** (Hardware Resource Participation, Hardware Resource)<br>22. **tailors** (Project Process, Standard Process): Project processes can be defined by tailoring standard processes.<br>23. **tailors** (Project Activity, Standard Activity): Proyect activities can be defined by tailoring standard activities.<br>24. **to perform** (Human Resource Allocation, Human Role)<br>25. **uses** (Activity Occurrence, Resource)<br><br>**Relation_without_name Role (TermX, TermY)**<br><br>26. **delegator** (Human Resource Allocation, Project Team): human resource allocation is a type of delegatum in UFO, in which the project team is the delegator.<br>27. **delegatee** (Human Resource Allocation, Human Resource): human resource allocation is a type of delegatum in UFO, in which the human resource is the delegatee.<br><br>**Axioms** are not available in [4]. However, we have considered the axioms formulated in the first version of SPO (available in the author's previous paper: "Falbo R. A. and Bertollo G.: A software process ontology as a common vocabulary about software processes. International Journal of Business Process Integration and Management, vol. 4, pp. 239–250, 2009"). This article is mentioned in [4], but without referring to its axioms. Probably, some of the 22 axioms specified in first-order logic for the SPO first version, do not fully apply to the new version of SPO [4] due |
| | to the harmonization with the foundational ontology named UFO and the addition of terms and semantic changes for some of them. Verifying the consistency of these axioms with [4] is beyond the scope of this comparative study. |
| **Ontological reuse** | <u>**Reuse of the Term from Foundational Ontology**</u> (UFO [4])<br>1. **Activity Occurrence** / UFO:: Action<br>2. **Atomic Activity Occurrence** / UFO:: Atomic Action<br>3. **Atomic Project Activity** / UFO::Intention (Internal Commitment)<br>4. **Atomic Standard Activity** / UFO:: Action Universal (plan) or UFO:: Atomic Action Universal<br>5. **Complex Activity Occurrence** / UFO:: Complex Action<br>6. **Complex Project Activity** / UFO::Complex Commitment<br>7. **Complex Standard Activity** / UFO::Action Universal (plan)<br>8. **General Project Process** / UFO::Intention (Internal Commitment)<br>9. **General Standard Process** / UFO::Complex Action Universal<br>10. **Hardware Equipment** / UFO:: Object<br>11. **Hardware Resource** / UFO:: Object<br>12. **Hardware Resource Participation** / UFO::Object Participation<br>13. **Hardware Type** / UFO::Object Kind<br>14. **Human Resource** / UFO::Human Agent<br>15. **Human Resource Allocation** / UFO:: Delegatum<br>16. **Human Resource Participation** / UFO:: Action Contribution<br>17. **Human Role** / UFO:: Social Role<br>18. **Job Plan** / UFO:: Normative Description<br>19. **Person** / UFO:: Human Agent<br>20. **Process Occurrence** / UFO:: Complex Action |
| | 21. **Project Activity /** UFO:: Intention (Internal Commitment)<br>22. **Project Process /** UFO:: Intention (Internal Commitment)<br>23. **Project Team /** UFO:: Collective Social Agent<br>24. **Project Team Allocation /** UFO:: Social Relator<br>25. **Resource /** UFO:: Object<br>26. **Scheduled Activity /** UFO::Appointment<br>27. **Scheduled Process /** UFO::Appointment<br>28. **Software Product /** UFO:: Object<br>29. **Software Product Type /** UFO:: Object Kind<br>30. **Software Resource /** UFO:: Object<br>31. **Software Resource Participation /** UFO::Object Participation<br>32. **Specific Project Process /** UFO:: Intention (Internal Commitment) or UFO::Complex Commitment<br>33. **Specific Standard Process /** UFO::Complex Action Universal<br>34. **Standard Activity /** UFO::Action Universal (plan)<br>35. **Standard Process /** UFO::Complex Action Universal<br>36. **Standard Process Definition Document /** UFO:: Plan Description<br><br>**Reuse of the Non-Taxonomic Relationship from Foundational Ontology** (UFO [4])<br><br>**Relation_name (TermX, TermY)**<br>1. **allocates** (Project Team Allocation, Human Resource)<br>2. **caused by** (Process Occurrence, Scheduled Process) / UFO::**caused by** (UFO::Action, UFO::Intention (Internal Commitment))<br>3. **caused by** (Activity Occurrence, Scheduled Activity) / UFO::**caused by** (UFO::Action, UFO::Intention (Internal Commitment))<br>4. **defines** (Job Plan, Human Role) / UFO::**defines** |
| | (UFO::Normative Description, UFO::Social Role)<br>5. **depends on** (Standard Activity, Standard Activity)<br>6. **depends on** (Activity Occurrence, Activity Occurrence)<br>7. **depends on** (Project Activity, Project Activity)<br>8. **describes** (Standard Process Definition Document, Standard Process) / UFO::**describes** (UFO::Plan Description, UFO::Complex Action Universal)<br>9. **existential dependence of** (Human Resource Allocation, Project Team Allocation)<br>10. **has** (Project Team, Project Team Allocation)<br>11. **is to be performed by** (Standard Activity, Human Role)<br>12. **is to be performed by** (Project Activity, Human Role)<br>13. **is to be performed using** (Standard Activity, Hardware Type)<br>14. **is to be performed using** (Standard Activity, Software Product Type)<br>15. **is to be performed using** (Project Activity, Hardware Type)<br>16. **is to be performed using** (Project Activity, Software Product Type)<br>17. **is to perform** (Project Team Allocation, Human Role)<br>18. **is to perform** (Human Resource Allocation, Scheduled Activity)<br>19. **participation of** (Human Resource Participation, Human Resource)<br>20. **participation of** (Software Resource Participation, Software Resource) / UFO::**participation of** (UFO::Object Participation, UFO::Object)<br>21. **participation of** (Hardware Resource Participation, Hardware Resource) / UFO::**participation of** (UFO::Object Participation, UFO::Object)<br>22. **tailors** (Project Process, Standard Process)<br>23. **tailors** (Project Activity, Standard Activity) |
| | 24. **to perform** (Human Resource Allocation, Human Role)<br>25. **uses** (Activity Occurrence, Resource)<br><br>**Relation_without_name Role (TermX, TermY)**<br>26. **delegator** (Human Resource Allocation, Project Team)<br>27. **delegatee** (Human Resource Allocation, Human Resource) |
| **International standard glossary used or referred** | **ISO/IEC 12207**, **ISO 9001:2000**, **ISO/IEC 15504**, **CMMI**, and **RUP**<br><br>Note that this information was collected from the author's previous paper: "Falbo R. A. and Bertollo G.: A software process ontology as a common vocabulary about software processes. International Journal of Business Process Integration and Management, vol. 4, pp. 239–250, 2009", which is referenced in [4]. |
| **Collector name** | Maria Fernanda Papa and Luis Olsina |
| **Article title** | Analyzing a Process Core Ontology and its Usefulness for Different Domains |
| **Author/s of the article** | Pablo Becker,<br>Fernanda Papa,<br>Guido Tebes, and<br>Luis Olsina |
| **Reference** | Analyzing a Process Core Ontology and its Usefulness for Different Domains, In Springer Nature book, CCIS 1439: Int'l Conference on the Quality of Information and Communication Technology, A. C. R. Paiva et al. (Eds.): QUATIC 2021, pp. 1–14, 2021 (see reference [3]) |
| **Name of the preselected ontology** | ProcessCO v1.2 |
| **Specified concepts used to describe the process core ontology** | <u>**Terms**</u>:<br>1. **Allocation** is an Assertion on Particulars, specifically, an Allotment-related Assertion that specifies the assignment of a Work Resource to a Work Entity.<br>2. **Allocation Model** represents an Artifact that specifies and models none or more Allocations of Work Resources.<br>3. **Activity** is a Work Entity that is formed by an interrelated set of sub-activities and Tasks.<br>4. **Agent** is a Work Resource assigned to a Work Entity to perform a Task in compliance with a Role.<br>5. **Artifact** is a tangible or intangible, versionable Work Product, which can be delivered.<br>6. **Automated Agent** is an Agent, in fact, a non-human Work Resource assigned to a Work Entity, which performs a Task in fulfillment of a Role.<br>7. **Condition** is a Constraint-related Assertion that specifies restrictions that must be satisfied or evaluated to true at the beginning (pre-condition) or ending (post-condition) of a Work Entity realization, in given project situations or events.<br>8. **Human Agent** is an Agent, in fact, a human Work Resource assigned to a Work Entity, which performs a Task in fulfillment of a Role. |
| | 9. **Method** is a Work Resource that encompasses the specific and particular way to perform the specified steps in the Work Entity description.<br><br>10. **Money** is a Work Resource that represents a medium of exchange accepted for the payment of goods, services and all kinds of obligations.<br><br>11. **Natural Product** is a Product Entity that is produced by natural processes.<br><br>12. **Outcome** is a Work Product that is intangible, storable and processable.<br><br>13. **Process Category** is a Thing Category (a universal) which has a Work Entity sub-Category.<br><br>14. **Process Model** represents an Artifact that specifies and models one or more related Process Perspectives.<br><br>15. **Process Perspective** is an Assertion on Particulars that specifies the functional, behavioral, informational, methodological, or organizational view for Work Entities and related concepts.<br><br>16. **Product Category** is a Thing Category (a universal) to which concrete Product Entities belong to.<br><br>17. **Product Entity** is a Thing (a particular) produced naturally or yielded artificially as a result of a Work Entity.<br><br>18. **Resource Category** is a Thing Category (a universal) to which concrete Resource Entities belong to.<br><br>19. **Resource Entity** is a Thing (a particular) that represents an available asset that can be intentionally used for or allocated to something as a means of help, support, or need in a particular event or situation.<br><br>20. **Role** is a Behavior-related Assertion that specifies a set of skills that an Agent must possess in order to perform a Work Entity.<br><br>21. **Service** is a Work Product that is intangible, non-storable and deliverable.<br><br>22. **Strategy** is a Work Resource that encompasses principles and integrated capabilities such as domain conceptual bases, the specification of process perspectives and methods for helping to achieve a project's goal purpose. |
| | 23. **Task** is an atomic, fine-grained Work Entity that cannot be decomposed.<br>24. **Time** is a Work Resource that represents a finite, non-storable, perishable and inexorable non-spatial continuum assigned to Work Entities (Processes, Activities and Tasks) when (re-)scheduling their duration in a project.<br>25. **Tool** is a Work Resource that represents an instrument facilitating the automation and execution of Method procedures and rules.<br>26. **Work Entity** is a Thing (a particular) that describes the work by means of consumed and produced Work Products, Conditions, and involved Roles.<br>27. **Work Entity sub-Category** is a Process's sub-category to which concrete Work Entities belong to.<br>28. **Work Process** is a coarse-grained Work Entity that is composed of an interrelated set of sub-processes and activities.<br>29. **Work Product** is a Product Entity that is consumed or produced by a Work Entity.<br>30. **Work Resource** is a Resource Entity that represents an available asset that can be allotted and assigned to Work Entities.<br><br>**Attributes (Properties)**:<br><br>1. Allocation (***name***=Label or name that identifies the Allocation of Work Resources, ***statement***= An unambiguous textual statement describing the Allocation of Work Resources)<br>2. Allocation Model (***specification***=The explicit and detailed representation or model of the Allocation in a given language)<br>3. Agent (***capabilities*** =Set of abilities that the Agent has as a performer)<br>4. Artifact (***state***= State in which the Artifact is, ***version***=Unique identifier, which indicates the level of evolution of the Artifact)<br>5. Condition (***specification***= Unambiguous specification of constraints, restrictions or circumstances that must be |

achieved or satisfied)

6. Method (***procedure***=Arranged set of instructions or operations, which specifies how the steps in a Task description must be performed, ***rules***=Set of principles, conditions, heuristics, axioms, etc. associated with the procedure, ***references***=Citation of bibliographical or URL resources, where authoritative and additional information for the Method can be consulted)
7. Outcome (***value***= Numerical or categorical result)
8. Process Model (***specification***= The explicit and detailed representation or model of the Work Entity perspective in a given language)
9. Process Perspective (***name***=Label or name that identifies the Process Perspective, ***statement***=An unambiguous textual statement describing the Process Perspective)
10. Product Entity (***name***=Label or name that identifies the Product Entity, ***description***=An unambiguous textual statement describing the Product Entity)
11. Resource Entity (***name***=Label or name that identifies the Resource Entity, ***description***= An unambiguous textual statement describing the Resource Entity)
12. Role (***name***=Label or name that identifies the Role, ***skills***= Set of capabilities, competencies and responsibilities of the Role)
13. Task (***Steps specification***=Specification of steps to be followed in order to achieve the Task objective)
14. Tool(***description***= An unambiguous textual statement describing the Tool, ***references***= Citation of bibliographical or URL resources, where authoritative and additional information for the Tool can be consulted)
15. Work Entity(***name***=Label or name that identifies the Work Entity, ***objective*** =Aim or end to be reached, ***description***=An unambiguous textual statement describing what to do for achieving the objective of the Work Entity, ***status***=State in which the Work Entity is, ***start date***= Date or instant of time when the Work Entity starts, ***end date***=Date or instant of time when the Work Entity ends)
16. Work Resource (***level***=Level to which the Work Resource is assigned)

| | |
|---|---|
| | **Relationships**:<br><br>**Taxonomic:**<br>**part_of (TermX_whole, TermY_part)**<br>1. part_of (Process Perspective, Process Model)<br>2. part_of (Process Category, Work Entity sub-Category)<br>3. part_of (Product Entity, Natural Product)<br>4. part_of (Product Entity, Work Product)<br>5. part_of (Work Process, Work Process)<br>6. part_of (Work Process, Activity)<br>7. part_of (Activity, Activity)<br>8. part_of (Activity, Task)<br>9. part_of (Allocation, Allocation Model)<br>**is_a (TermX_type, TermY_subtype)**<br>10. is_a (Work Entity, Work Process)<br>11. is_a (Work Entity, Activity)<br>12. is_a (Work Entity, Task)<br>13. is_a (Product Entity, Work Product)<br>14. is_a (Product Entity, Natural Product)<br>15. is_a (Work Product, Outcome)<br>16. is_a (Work Product, Artifact)<br>17. is_a (Work Product, Service)<br>18. is_a (Work Resource, Method)<br>19. is_a (Work Resource, Tool)<br>20. is_a (Work Resource, Money)<br>21. is_a (Work Resource, Time)<br>22. is_a (Work Resource, Strategy)<br>23. is_a (Work Resource, Agent)<br>24. is_a (Agent, Human Agent) |
| | 25. is_a (Agent, Automated Agent)<br>26. is_a (Resource Entity, Work Resource)<br>**Non-Taxonomic:**<br>**Relation_name (TermX, TermY)**<br>1. **consumes** (Work Entity, Product Entity): In order to achieve its objective, a Work Entity consumes one or more Product Entities.<br>2. **deals with** (Allocation, Work Resource): An Allocation deals with one or more Work Resources.<br>3. **deals with work entity** (Process Perspective, Work Entities): A Process Perspective deals with one or more Work Entities.<br>4. **involves** (Work Entity, Role): A Work Entity involves one or more Roles. In turn, a Role may participate in one or more Work Entities.<br>5. **is applicable** (Method, Task): A Method is applicable to the description of a Task. In turn, for a Task's description one or several Methods can be applied.<br>6. **is assigned to** (Allocation, Work Entity): A scheduled Allocation of Work Resources is assigned to Work Entities for their enactment.<br>7. **is played by** (Role, Agent): A Role is played by one or several Agents. In turn, an Agent plays one or more Roles.<br>8. **is related with** (Product Entity, Product Entity): A Product Entity is related with none or several Product Entities.<br>9. **ir required by** (Tool, Method): A Tool is required by none or several Methods.<br>10. **performs** (Agent, Task): An Agent performs one or more assigned Tasks. In turn, a Task is performed by one or more Agents.<br>11. **pertains to category** (Work Entity, Work Entity sub-Category): Work Entities pertain to a Work Entity sub-Category.<br>12. **pertains to product category** (Product Entity, Product Category): Product Entities pertain to a Product |
| | Category.<br><br>13. **pertains to resource category** (Resource Entity, Resource Category): Work Resources pertain to a Resource Category.<br><br>14. **produces** (Work Entity, Work Product): A Work Entity produces (modifies, creates) one or more Work Products.<br><br>15. **sets postcondition** (Work Entity, Condition): A Work Entity may have associated Conditions, which must be accomplished at the end of its realization to be considered finished.<br><br>16. **sets precondition** (Work Entity, Condition): A Work Entity may have associated Conditions, which must be accomplished before its initiation.<br><br>17. **relates** (Process Perspective, Process Perspective): A Process Perspective relates with none or several Process Perspectives.<br><br>18. **uses**(Agent, Work Resource): An Agent uses one or more Work Resources to perform a Task.<br><br>**Axioms** are not available. |
| **Ontological reuse** | **Reuse of the Term from Foundational Ontology** (ThingFO [10, 11])<br><br>1. **Allocation /** ThingFO :: Allotment-related Assertion<br><br>2. **Allocation Model /** ThingFO :: Thing<br><br>3. **Activity /** ThingFO :: Thing<br><br>4. **Agent /** ThingFO :: Thing<br><br>5. **Artifact /** ThingFO :: Thing<br><br>6. **Automated Agent /** ThingFO :: Thing<br><br>7. **Condition /** ThingFO :: Constraint-related Assertion<br><br>8. **Human Agent /** ThingFO :: Thing<br><br>9. **Method /** ThingFO :: Thing<br><br>10. **Money /** ThingFO :: Thing<br><br>11. **Natural Product /** ThingFO :: Thing |

12. **Outcome** / ThingFO :: Thing
13. **Process Category** / ThingFO :: Thing Category
14. **Process Model** / ThingFO :: Thing
15. **Process Perspective** / ThingFO :: Assertion on Particulars
16. **Product Category** / ThingFO :: Thing Category.
17. **Product Entity** / ThingFO :: Thing
18. **Resource Category** / ThingFO :: Thing Category
19. **Resource Entity** / ThingFO :: Thing
20. **Role** / ThingFO :: Behavior-related Assertion
21. **Service** / ThingFO :: Thing
22. **Strategy** / indirectly from ThingFO :: Thing
23. **Task** / ThingFO :: Thing
24. **Time** / ThingFO :: Thing
25. **Tool** / ThingFO :: Thing
26. **Work Entity** / ThingFO :: Thing
27. **Work Entity sub-Category** / ThingFO :: Thing Category
28. **Work Process** / ThingFO :: Thing
29. **Work Product** / ThingFO :: Thing
30. **Work Resource** / ThingFO :: Thing

**<u>Reuse of the Non-Taxonomic Relationship from Foundational Ontology</u>** (ThingFO [10, 11])

**Relation_name (TermX, TermY)**

1. **consumes** (Work Entity, Product Entity) / ThingFO:: **interacts with other**( ThingFO::(Power of) Thing, ThingFO::Thing)
2. **deals with** (Allocation, Work Resource) / ThingFO::**deals with particulars**( ThingFO::Assertion on Particulars, ThingFO::Thing)
3. **deals with work entity** (Process Perspective, Work Entities) / ThingFO::**deals with particulars**(

| | |
|---|---|
| | ThingFO::Assertion on Particulars, ThingFO::Thing)<br>4. **involves** (Work Entity, Role) / ThingFO:: **defines**(ThingFO::Thing, ThingFO::Assertion)<br>5. **is applicable** (Method, Task) / ThingFO:: **relates with**(ThingFO::Thing, ThingFO::Thing)<br>6. **is assigned to** (Allocation, Work Entity) / ThingFO::**deals with particulars**(ThingFO::Assertion on Particulars, ThingFO::Thing)<br>7. **is played by** (Role, Agent) / ThingFO::**deals with particulars**(ThingFO::Assertion on Particulars, ThingFO::Thing)<br>8. **is related with** (Product Entity, Product Entity) / ThingFO:: **relates with**(ThingFO::Thing, ThingFO::Thing)<br>9. **ir required by** (Tool, Method) / **ThingFO::interacts with other**(**ThingFO::**(Power of) Thing, **ThingFO::**Thing)<br>10. **performs** (Agent, Task) / **ThingFO::interacts with other**(**ThingFO::**(Power of) Thing, **ThingFO::**Thing)<br>11. **pertains to category** (Work Entity, Work Entity sub-Category) / ThingFO::**belongs to**(**ThingFO::**Thing, **ThingFO::**Thing Category)<br>12. **pertains to product category** (Product Entity, Product Category) / ThingFO::**belongs to**(**ThingFO::**Thing, **ThingFO::**Thing Category)<br>13. **pertains to resource category** (Resource Entity, Resource Category) / ThingFO::**belongs to**(**ThingFO::**Thing, **ThingFO::**Thing Category)<br>14. **produces** (Work Entity, Work Product) / ThingFO:: **interacts with other**( ThingFO::(Power of) Thing, ThingFO::Thing)<br>15. **sets postcondition** (Work Entity, Condition) / ThingFO:: **defines**(ThingFO::Thing, ThingFO:: Assertion)<br>16. **sets precondition** (Work Entity, Condition) / ThingFO:: **defines**(ThingFO::Thing, ThingFO:: Assertion)<br>17. **relates** (Process Perspective , Process Perspective) / |
| | ThingFO::**relates with**(ThingFO::Assertion on Particulars, ThingFO::Assertion on Particulars)<br>18. **uses**(Agent, Work Resource) / ThingFO:: **interacts with other**( ThingFO::(Power of) Thing, ThingFO::Thing) |
| **International standard glossary used or referred** | **ISO/IEC 12207**, **SPEM**, and **CMMI**<br><br>Note that this information was collected from the author's previous paper: "Becker P., Papa F., Olsina L.: Process Ontology Specification for Enhancing the Process Compliance of a Measurement and Evaluation Strategy. In CLEI Electronic Journal, 18(1), pp. 1–26, https://doi.org/10.19153/cleiej.18.1.2, 2015", which is referenced in [3]. |

# Appendix V: Base and Derived Measure Values for SPO and ProcessCO v1.2

*This Appendix documents the measure values produced by measurement using direct and indirect metrics that are specified in Appendix II. Note that the # symbol represents 'amount of' and only applies to direct metrics, while the % symbol represents 'percentage' which is a proportion that only applies to indirect metrics.*

| | | ***SPO*** | ***ProcessCO*** | |
|---|---|---|---|---|
| **%DT** | | 80.56 | 100.00 | **%DT = (#DT / #TT) * 100** |
| | #DT | 29 | 30 | |
| | #TT | 36 | 30 | |

| | | ***SPO*** | ***ProcessCO*** | |
|---|---|---|---|---|
| **%DP** | | 0.00 | 100.00 | **%DP = (#DP / #TP) * 100;**<br>If **#TP = 0**, then **%DP = 0** |
| | #DP | 0 | 30 | |
| | #TP | 0 | 30 | |

| | | ***SPO*** | ***ProcessCO*** | |
|---|---|---|---|---|
| **%SA** | | 100.00 | 0.00 | **%SA = (#SA / #TA) * 100**;<br>If **#TA = 0**, then **%SA = 0** |
| | #SA | 22 | 0 | |
| | #TA | 22 | 0 | |

| | | ***SPO*** | ***ProcessCO*** | |
|---|---|---|---|---|
| **%DNTR** | | 40.74 | 100.00 | **%DNTR = (#DNTR / #TNTR) * 100**;<br>If **#TNTR = 0**, then **%DNTR = 0** |
| | #DNTR | 11 | 18 | |
| | #TNTR | 27 | 18 | |

| | | ***SPO*** | ***ProcessCO*** | |
|---|---|---|---|---|
| **%BNTR** | | 50.94 | 40.91 | **%BNTR = (#TNTR / #TR) * 100** |
| | #TNTR | 27 | 18 | |
| | #TR | 53 | 44 | |

| | | ***SPO*** | ***ProcessCO*** | |
|---|---|---|---|---|
| **%STFO** | | 100.00 | 100.00 | **%STFO = ((#STDFO + #STIFO) / #TT) * 100** |
| | #STDFO | 36 | 30 | |
| | #STIFO | 0 | 0 | Not considered, computing it in #STDFO |
| | #TT | 36 | 30 | |

| | | ***SPO*** | ***ProcessCO*** | |
|---|---|---|---|---|
| **%SNTRFO** | | 22.22 | 100.00 | **%SNTRFO = (#SNTRFO / #TNTR) * 100**<br>If **#TNTR** = 0, then **%SNTRFO** = 0 |
| | #SNTRFO | 6 | 18 | |
| | #TNTR | 27 | 18 | |

| | ***SPO*** | ***ProcessCO*** |
|---|---|---|
| **#UISG** | 5 | 3 |

# Appendix VI: Elementary Indicator Values for SPO and ProcessCO v1.2

*This Appendix shows the elementary indicator values (expressed in %) calculated from the measured values (Appendix V) and using the elementary function of the indicators that are specified in Appendix III.*

| | ***SPO*** | ***ProcessCO*** | |
|---|---|---|---|
| **PL_DTA** | 80.56 | 100 | PL_DTA = %DT |
| **PL_DPA** | 0 | 100 | PL_DPA = %DP |
| PL_FSAA | 100 | 0 | PL_FSAA = %SA |
| **PL_DNTRA** | 40.74 | 100 | PL_DNTRA = %DNTR |
| **PL_BNTTRA** | 98.58 | 86.36 | $\mathbf{PL_BNTTRA}(x) = \begin{cases} 6x; & 0 \le x < 10 \quad (1) \\ \frac{5}{6}x + \frac{155}{3}; & 10 \le x < 40 \quad (2) \\ \frac{3}{2}x + 25; & 40 \le x \le 50 \quad (3) \\ -\frac{3}{2}x + 175; & 50 < x \le 60 \quad (4) \\ -\frac{5}{6}x + 135; & 60 < x \le 90 \quad (5) \\ -6x + 600; & 90 < x \le 100 \quad (6) \end{cases}$ 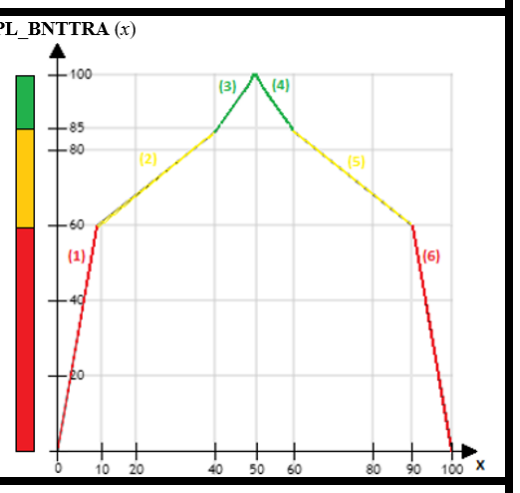  |
| **P_LRTFO** | 100 | 100 | $\mathbf{P_LRTFO} = \begin{cases} 100; & \%\mathbf{STFO} = 100 \\ 85; & 95 \le \%\mathbf{STFO} < 100 \\ 60; & 70 \le \%\mathbf{STFO} < 95 \\ 0; & 0 \le \%\mathbf{STFO} < 70 \end{cases}$ |
| **P_LRNTRFO** | 20 | 100 | $\mathbf{P_LRNTRFO} = \begin{cases} 100; & \%\mathbf{SNTRFO} = 100 \\ 85; & 95 \le \%\mathbf{SNTRFO} < 100 \\ 60; & 70 \le \%\mathbf{SNTRFO} < 95 \\ 20; & 20 \le \%\mathbf{SNTRFO} < 70 \\ 0; & 0 \le \%\mathbf{SNTRFO} < 20 \end{cases}$ |
| **P_LUISG** | 100 | 100 | $\mathbf{P_LUISG} = \begin{cases} 0 \; if \; \#\mathbf{UISG} = 0 \\ 75 \; if \; \#\mathbf{UISG} = 1 \\ 100 \; if \; \#\mathbf{UISG} \ge 2 \end{cases}$ |

# Appendix VII: Derived Indicator Values for SPO and ProcessCO v1.2

*This Appendix shows the yielded values of derived indicators (expressed in %) that evaluate the quality model of Appendix I. The partial and global values are calculated from the values of elementary indicators (Appendix VI), and using the LSP aggregation function [5], which supports weights and operators of simultaneity or conjunction (operators C), of replaceability or disjunction (operators D), and of independence of the inputs to produce an output (operator A).*

| | **Weights** | **Op.** | ***SPO*** | ***ProcessCO*** |
|---|---|---|---|---|
| 1 Ontological Internal Quality | | C+ | **64.81** | **87.82** |
| 1.1 Ontological Structural Quality | 0.6 | A | **61.12** | **82.52** |
| *1.1.1 Defined Terms Availability* | 0.3 | | ***80.56*** | ***100.00*** |
| *1.1.2 Defined Properties Availability* | 0.25 | | ***0.00*** | ***100.00*** |
| *1.1.3 Formally Specified Axioms Availability* | 0.15 | | ***100.00*** | ***0.00*** |
| 1.1.4 Balanced Relationships Availability | 0.3 | C— | **73.17** | **91.73** |
| *1.1.4.1 Defined Non-Taxonomic Relationships Availability* | 0.4 | | ***40.74*** | ***100.00*** |
| *1.1.4.2 Balanced Non-Taxonomic / Taxonomic Relationships Availability* | 0.6 | | ***98.58*** | ***86.36*** |
| 1.2 Ontological Quality of Reuse and Compliance | 0.4 | C— | **73.06** | **100.00** |
| 1.2.1 Ontological Reuse Quality | 0.7 | C— | **62.52** | **100.00** |
| *1.2.1.1 Level of Reuse of Terms from Foundational Ontology* | 0.6 | | ***100.00*** | ***100.00*** |
| *1.2.1.2 Level of Reuse of Non-Taxonomic Relationships from Foundational Ontology* | 0.4 | | ***20.00*** | ***100.00*** |
| *1.2.2 Level of Use of International Standard Glossaries* | 0.3 | | ***100.00*** | ***100.00*** |

# Appendix VIII: Derived Indicator Values after change and revaluation performed for ProcessCO

*This Appendix shows the yielded values of derived indicators (expressed in %) that evaluate the quality model of Appendix I only for ProcessCO and its improvement. The partial and global values are calculated using the same LSP aggregation function with the same weights and operators for both the comparative study shown in Appendix VII and for the revaluation of the improved entity. For the change made over ProcessCO v1.2, attribute 1.1.3 was basically considered since it fell into the red acceptability level, which implied the addition of six axioms specified in first-order logic. As a result of this change, the new version (v1.3) of ProcessCO was issued, which can be accessed in [2]. In the following table the reader can appreciate the increase achieved in Ontological Internal Quality for ProcessCO v1.3.*

| | | | ***ProcessCO*** | |
|---|---|---|---|---|
| | **Weights** | **Op.** | ***v1.2*** | ***v1.3*** |
| 1 Ontological Internal Quality | | C+ | **87.82** | **98.48** |
| 1.1 Ontological Structural Quality | 0.6 | A | **82.52** | **97.52** |
| *1.1.1 Defined Terms Availability* | 0.3 | | ***100.00*** | ***100.00*** |
| *1.1.2 Defined Properties Availability* | 0.25 | | ***100.00*** | ***100.00*** |
| *1.1.3 Formally Specified Axioms Availability* | 0.15 | | ***0.00*** | ***100.00*** |
| 1.1.4 Balanced Relationships Availability | 0.3 | C— | **91.73** | **91.73** |
| *1.1.4.1 Defined Non-Taxonomic Relationships Availability* | 0.4 | | ***100.00*** | ***100.00*** |
| *1.1.4.2 Balanced Non-Taxonomic / Taxonomic Relationships Availability* | 0.6 | | ***86.36*** | ***86.36*** |
| 1.2 Ontological Quality of Reuse and Compliance | 0.4 | C— | **100.00** | **100.00** |
| 1.2.1 Ontological Reuse Quality | 0.7 | C— | **100.00** | **100.00** |
| *1.2.1.1 Level of Reuse of Terms from Foundational Ontology* | 0.6 | | ***100.00*** | ***100.00*** |
| *1.2.1.2 Level of Reuse of Non-Taxonomic Relationships from Foundational Ontology* | 0.4 | | ***100.00*** | ***100.00*** |
| *1.2.2 Level of Use of International Standard Glossaries* | 0.3 | | ***100.00*** | ***100.00*** |